\documentclass[12pt]{article}

\usepackage[dvipdfmx]{graphicx}
\usepackage{subcaption}
\usepackage{listings,jvlisting}
\usepackage{amsmath,amssymb}
\usepackage{bm}
\usepackage{graphicx, color}
\usepackage{wrapfig}
\usepackage{cite}
\usepackage{braket}
\begin{document}
\title{
\begin{flushright}
\ \\*[-80pt]
\begin{minipage}{0.2\linewidth}
\normalsize
%arXiv:YYMM.NNNN \\
HUPD-2210 \\*[50pt]
\end{minipage}
\end{flushright}
{\Large \bf
Non-SUSY Lepton Flavor Model with 3HDM
\\*[20pt]}}

\author{
\centerline{
Yukimura~Izawa $^{1}$\footnote{izawa-yukimura@hiroshima-u.ac.jp},~
%~ $^{1}$\footnote{},} \\*[5pt]
%\centerline{
Yusuke~Shimizu $^{1,2}$\footnote{yu-shimizu@hiroshima-u.ac.jp},
~Hironori~Takei $^{1}$\footnote{t-hironori@hiroshima-u.ac.jp}}
\\*[20pt]
\centerline{
\begin{minipage}{\linewidth}
\begin{center}
$^1${\it \normalsize
Physics Program, Graduate School of Advanced Science \\ and Engineering,~Hiroshima~University, \\
Higashi-Hiroshima~739-8526,~Japan \\*[5pt]
$^2${\it \normalsize
Core of Research for the Energetic Universe, Hiroshima University, \\
Higashi-Hiroshima 739-8526, Japan}
}
\end{center}
\end{minipage}}
\\*[50pt]}

\date{
\centerline{\small \bf Abstract}
\begin{minipage}{0.9\linewidth}
\medskip
\medskip
\small
We propose a simple non-supersymmetric lepton flavor model with $A_4$ symmetry.
The $A_4$ group is a minimal one which includes triplet irreducible representation.
We introduce three Higgs doublets which are assigned as triplet of the $A_4$ symmetry.
It is natural that there are three generations of the Higgs fields as same as 
the standard model fermions.
We analyse the potential and we get the vacuum expectation values 
for the local minimum.
In the vacuum expectation values, we obtain the charged lepton, Dirac neutrino, 
and right-handed Majorana neutrino mass matrices. By using type-I seesaw mechanism, 
we get the left-handed Majorana neutrino mass matrix.
In the NuFIT 5.1 data, we predict the Dirac CP phase and the Majorana phases for the only 
inverted neutrino mass hierarchy. 
Especially, the Dirac CP phase and lepton mixing angle $\theta _{23}$ are strongly correlated. 
If the $\theta _{23}$ is more precise measured, 
the Dirac CP phase is more precise predicted, and vice versa. 
We also predict the effective mass for neutrino-less double beta decay 
$m_{ee}\simeq 47.1~[\mathrm{meV}]$ and the lightest neutrino mass  
$m_3\simeq 0.789$~-~$1.43~[\mathrm{meV}]$.
It is testable for our model in the near future neutrino experiments.
\end{minipage}
}

\begin{titlepage}
\maketitle
\thispagestyle{empty}
\end{titlepage}
\newpage
%------------------------------------------------------------------------------%
%--------------------------------   Introduction   --------------------------------%
%------------------------------------------------------------------------------%
\section{Introduction}
The standard model (SM) is the successful one with the discovery of the Higgs boson. 
In the particle physics, the gauge theory is applied and tested 
by the electroweak precision measurements for the SM.
However there are still mysterious puzzles, e.g. the origin of the generations 
which are differences of the mixing angles and masses for quark and lepton sectors.
Actually, the Yukawa couplings are completely free parameters so that 
the mixing angles and masses cannot be predicted in the SM. 
In addition, the neutrinos are massless for renormalizable operators.
One of the attractive phenomena to solve the puzzles is the neutrino oscillation
which provides us the useful informations such as three lepton mixing angles and 
two neutrino mass squared differences which mean that the neutrinos have non-zero masses.
The T2K and NO$\nu $A experiments have confirmed the neutrino oscillation 
in the $\nu _\mu \to \nu _e$ appearance 
events~\cite{T2K:2013bqz,T2K:2013ppw,NOvA:2016kwd},
%\cite{T2K:2013bqz}--\cite{NOvA:2016kwd}, 
which are one of the clues of the new physics beyond the SM such as
the Dirac CP violating phase for the lepton sector by combining the data of the 
reactor neutrino  experiments~\cite{DayaBay:2012fng,RENO:2012mkc}.
The KamLAND-Zen~\cite{KamLAND-Zen:2016pfg,KamLAND-Zen:2022tow}, 
GERDA\cite{GERDA:2013vls,GERDA:2018pmc}, and 
CUORE~\cite{CUORE:2017tlq,CUORE:2019yfd} 
experiments also provide us the significant informations
which are whether the neutrinos are Dirac or Majorana particles, the lepton number violation,
and Majorana phases if the neutrinos are Majorana particles.
Thus the neutrino oscillation experiments go into a new phase of the precise determinations
of the lepton mixing angles, the neutrino mass squared differences, and the CP violating phases.

The SM particles obey the gauge theory. After spontaneous symmetry breaking (SSB),
the gauge bosons and fermions get the masses through the Higgs mechanism.
However the Yukawa couplings cannot be controlled by the gauge symmetry, 
the Yukawa couplings
are completely free parameters in the SM.
The flavor symmetry can apply to the generations. 
The Froggatt-Nielsen mechanism which was proposed by C.~D.~Froggatt and H.~B.~Nielsen
are introduced global $U(1)_\text{FN}$ symmetry~\cite{Froggatt:1978nt}. 
Thanks for the $U(1)_\text{FN}$ symmetry, it is natural to explain the fermion mass hierarchies. 
On the other hand, the non-Abelian discrete symmetry 
(See for the review~\cite{Altarelli:2010gt}-\cite{Kobayashi:2022moq}.) 
can naturally explain the lepton mixing angles so-called 
``tri-bimaximal mixing (TBM)''~\cite{Harrison:2002er,Harrison:2002kp}
before the reactor experiments reported the non-zero reactor angle 
$\theta _{13}$~\cite{DayaBay:2012fng,RENO:2012mkc}.
Actually, many authors have studied the breaking or deviation from the 
TBM~\cite{Xing:2002sw}-\cite{King:2013eh}
or other patterns of the lepton mixing angles,
e.g. tri-bimaximal-Cabibbo mixing~\cite{King:2012vj,Shimizu:2012ry}.
One of the successful flavor models was proposed 
by G.~Altarelli and F.~Feruglio~\cite{Altarelli:2005yp,Altarelli:2005yx}.
The Altarelli and Feruglio (AF) model show the TBM by using 
the non-Abelian discrete symmetry $A_4$.
They introduced two SM gauge singlet scalar fields so-called ``flavons'' and taking
the vacuum expectation value (VEV) alignments of the $A_4$ triplets as 
$(1,0,0)$ and $(1,1,1)$,
which is naturally explained that the charged lepton is diagonal and neutrino mixing is TBM, respectively.
However the corrected VEV alignments cannot be driven from the potential analysis. 
Then, they applied to the supersymmetry (SUSY) and introduced the ``driving'' fields.
There are so many scalar fields in addition to the no evidence of the SUSY particles 
for the accelerator experiments, e.g. Large Hadron Collider experiment.

In this paper, we propose a simple non-SUSY lepton flavor model with $A_4$ symmetry.
The $A_4$ group is a minimal one which includes triplet irreducible representation.
We introduce three Higgs doublets which is assigned as triplet of the $A_4$ symmetry.
It is natural that there are three generations as same as the SM fermions.
We analyse the potential and we get the VEV for the local minimum 
in the three Higgs doublet model (3HDM)~\cite{Darvishi:2021txa} 
with $A_4$ symmetry~\cite{Lavoura:2007dw}-\cite{Carrolo:2022oyg}. 
The left-handed lepton doublets are assigned to triplet and the right-handed 
charged leptons are assigned to different singlets of the $A_4$ symmetry, respectively.
We introduce the right-handed Majorana neutrinos which are assigned to triplet of the $A_4$ symmetry. 
In our model, the right-handed Majorana neutrino mass matrix has a simple flavor structure. 
On the other hand, the Dirac neutrino mass matrix has symmetric and anti-symmetric 
Yukawa couplings for $A_4$ symmetry.
By using the type-I seesaw 
mechanism~\cite{Minkowski:1977sc,Yanagida,Gell-Mann:1979vob,Mohapatra:1979ia,Schechter:1980gr}, 
we obtain the left-handed Majorana neutrino mass matrix.
After diagonalizing the charged lepton and left-handed Majorana neutrino mass matrices, 
we get the lepton mixing matrix which is Pontecorvo-Maki-Nakagawa-Sakata (PMNS) 
one~\cite{Maki:1962mu,Pontecorvo:1967fh}.
In the numerical analysis, we use the NuFIT 5.1 
data~\cite{Esteban:2020cvm,Gonzalez-Garcia:2021dve}.
%%%%%%%%%   20240910 Takei added %%%%%%%%%%%%%%%%
We find that only inverted ordering is acceptable and we cannot find the solutions 
for normal ordering in the neutrino mass hierarchy. 
%normal ordering is rejected.
We obtain relevant relations for mixing angles and the effective mass for the neutrino-less 
double beta ($0\nu \beta \beta $) decay as a function of the lightest neutrino mass.
Especially, the Dirac CP phase and lepton mixing angle $\theta _{23}$ are strongly correlated. 
If the neutrinos are Majorana particles, the effective mass for the 
neutrino-less double beta decay and the lightest neutrino mass are also predicted in our model.

This paper is organized as follows. In Section~\ref{sec:3HDM}, 
we briefly introduce the $A_4$ group. Next, we analyse the potential with $A_4$ symmetry.
In Section~\ref{sec:model}, we present the $A_4$ flavor model and study mass matrices.
In Section~\ref{sec:Numerical}, we show the numerical analysis of our model.  
Section~\ref{sec:Summary} is devoted to a summary and discussions.
%In Appendix~\ref{sec:app_Higgs}, we show the Higgs mass matrices in the general 3HDM.
We show the relevant multiplication rule of the $A_4$ group 
in Appendix~\ref{sec:multiplication-rule}.

%------------------------------------------------------------------------------%
%---------------------------------3HDM----------------------------------------%%------------------------------------------------------------------------------%
\section{Potential analysis in the 3HDM with $A_4$ symmetry}
\label{sec:3HDM}
In this section, we discuss the 3HDM. First, we briefly introduce 
the $A_4$ group. Next, we analyse the potential with $A_4$ symmetry in the 3HDM, where we assign the three Higgs doublets as 
the triplet of the $A_4$ symmetry. 
It is natural that the Higgs fields are three generations as same as the SM fermions.
Let us analyse the potential in the 3HDM with $A_4$ symmetry. 
The $A_4$ group which is a minimal including triplet irreducible representation 
is the symmetric group of a tetrahedron or even permutation of four elements.
%a minimal one which includes triplet irreducible representation. 
There are twelve elements and four irreducible representations such as 
three different singlets $\bf{1}$, ${\bf 1'}$, ${\bf 1''}$ and triplet ${\bf 3}$ 
in the $A_4$ group, respectively.
Also the $A_4$ can be defined as the group generated by two elements $S$ and $T$ which 
satisfy the following algebraic relations as
\begin{equation}
S^2=T^3=(ST)^3={\bf 1}.
\end{equation}
These generators are represented by
\begin{align}
\label{eq:one-dimensional_rep}
{\bf 1}:& \quad S=1,\quad T=1, \nonumber \\
{\bf 1'}:& \quad S=1,\quad T=e^{\frac{4\pi i}{3}}\equiv \omega ^2, \\
{\bf 1''}:& \quad S=1,\quad T=e^{\frac{2\pi i}{3}}\equiv \omega , \nonumber 
\end{align}
on the one-dimensional representations.
These generators are also represented by 
\begin{equation}
\label{eq:three-dimensional_rep}
{\bf 3}:\quad S=\frac{1}{3}
\begin{pmatrix}
-1 & 2 & 2 \\
2 & -1 & 2 \\
2 & 2 & -1 
\end{pmatrix},\quad T=
\begin{pmatrix}
1 & 0 & 0 \\
0 & \omega ^2 & 0 \\
0 & 0 & \omega 
\end{pmatrix}, 
\end{equation}
on the three-dimensional representation. 
In these bases Eqs.~\eqref{eq:one-dimensional_rep} and \eqref{eq:three-dimensional_rep}, 
we can make the character table and obtain the multiplication rules of the $A_4$ group. 
The relevant multiplication rule is shown in Appendix~\ref{sec:multiplication-rule}. 

We introduce three Higgs doublets $\phi _1,\phi _2,\phi _3$ 
which are assigned as triplet $\Phi $:
%Suppose that a scalar field $\phi$ which is an $SU(2)_L$ doublet appearing in the SM 
%is a triplet as
\begin{equation}
	\Phi=(\phi_1,\phi_2,\phi_3),
\end{equation}
of the $A_4$ symmetry.
On the other hand, the complex conjugate of the $\Phi$ is 
considered by the conjugate of the generator $T$ of Eq.~\eqref{eq:three-dimensional_rep} as
\begin{equation}
T^*=
\begin{pmatrix}
1 & 0 & 0 \\
0 & \omega & 0 \\
0 & 0 & \omega ^2
\end{pmatrix}.
\label{eq:complex_generator_T}
\end{equation}
Then, the complex conjugate of the $\Phi $ is given by 
\begin{equation}
\Phi ^*=(\phi _1^*,\phi _3^*,\phi _2^*),
\end{equation}
and the multiplication rule of $A_4$ is kept as Eq.~\eqref{eq:multiplication-rule} 
in Appendix~\ref{sec:multiplication-rule}.
In our model, the Higgs potential is written as %of the SM 
\begin{equation}
	V=-\mu^2 \Phi^{\dag} \Phi +\lambda (\Phi^{\dag} \Phi)^2.
\end{equation}
This potential is invariant for hte $SU(2)_L \times U(1)_Y$ of the SM and $A_4$ symmetry.
By using the multiplication rule of $A_4$ in Appendix~\ref{sec:multiplication-rule}, we obtain the Higgs potential as follows:
\begin{align} \label{eq:A4_potential}
	V=&-\mu^2(|\phi_1|^2+|\phi_2|^2+|\phi_3|^2)\nonumber \\
	&+\lambda_1|\phi_1^2+2\phi_2\phi_3|^2+\lambda_2|\phi_2^2+2\phi_3\phi_1|^2+\lambda_3|\phi_3^2+2\phi_1\phi_2|^2\notag \\
	&+\lambda_4\Big[|\phi_1^2-\phi_2\phi_3|^2+|\phi_2^2-\phi_3\phi_1|^2
	+|\phi_3^2-\phi_1\phi_2|^2\Big],
\end{align}
where we rewrite the coupling $4\lambda_4/9$ as $\lambda_4$ in our convention.
%The VEVs of these Higgs fields are as follows:
%\begin{equation}
%	\langle\phi_1\rangle=v_1, \quad \langle\phi_2\rangle=v_2, \quad \langle\phi_3\rangle=v_3,
%\end{equation}
%where $v_1, ~v_2, ~v_3$ are real and we consider $\phi_1,\phi_2,\phi_3$ as three scalar fieldsas for simplicity. 
%When the Higgs fields have these values, the potential takes the minimum value, so the following holds:
We consider the potential minimum conditions as
\begin{align} 
	\left(\frac{\partial V}{\partial \phi_1}\right)_{\phi_1=\langle\phi_1\rangle,\phi_2=\langle\phi_2\rangle,\phi_3=\langle\phi_3\rangle}&=0,\\
	\left(\frac{\partial V}{\partial \phi_2}\right)_{\phi_1=\langle\phi_1\rangle,\phi_2=\langle\phi_2\rangle,\phi_3=\langle\phi_3\rangle}&=0,\\
	\left(\frac{\partial V}{\partial \phi_3}\right)_{\phi_1=\langle\phi_1\rangle,\phi_2=\langle\phi_2\rangle,\phi_3=\langle\phi_3\rangle}&=0.
\label{eq:minimum_condition}
\end{align}
From Eqs.~\eqref{eq:A4_potential}-\eqref{eq:minimum_condition} we obtain the following conditions:
\begin{align} 
	0=&-2\mu^2 v_1+4\lambda_1 v_1(v_1^2+2v_2v_3)+4\lambda_2 v_3(v_2^2+2v_3v_1)+4\lambda_3 v_2(v_3^2+2v_1v_2) \notag \\ \label{eq:vacuum_cond1}
	&+4\lambda_4\Big[v_1(v_1^2-v_2v_3)-\frac{1}{2}v_2(v_3^2-v_1v_2)-\frac{1}{2}v_3(v_2^2-v_3v_1)\Big], \\
	0=&-2\mu^2 v_2+4\lambda_1 v_3(v_1^2+2v_2v_3)+4\lambda_2 v_2(v_2^2+2v_3v_1)+4\lambda_3 v_1(v_3^2+2v_1v_2) \notag \\ \label{eq:vacuum_cond2}
	&+4\lambda_4\Big[v_2(v_2^2-v_3v_1)-\frac{1}{2}v_1(v_3^2-v_1v_2)-\frac{1}{2}v_3(v_1^2-v_2v_3)\Big], \\
	0=&-2\mu^2 v_3+4\lambda_1 v_2(v_1^2+2v_2v_3)+4\lambda_2 v_1(v_2^2+2v_3v_1)+4\lambda_3 v_3(v_3^2+2v_1v_2) \notag \\ \label{eq:vacuum_cond3}
	&+4\lambda_4\Big[v_3(v_3^2-v_1v_2)-\frac{1}{2}v_1(v_2^2-v_3v_1)-\frac{1}{2}v_2(v_1^2-v_2v_3)\Big]. 
\end{align}

%We find the solutions $(0,0,0)$, $(1,0,0)$, and general solution $(v_1,v_2,v_3)$, $v_1\neq v_2\neq v_3$. 
%The trivial solution and $(1,0,0)$ case cannot be realized the current experimental data. 
%In the general solution, we cannot predict the physical quantity because there are many free parameters. 
%Therefore, we should use one of the solutions which we can predict the Dirac CP phase, Majorana phases, and effective mass for the $0\nu \beta \beta $ decay.
%Then, we analyse the potential more detail. 
We sum the conditions Eqs.~\eqref{eq:vacuum_cond1}-\eqref{eq:vacuum_cond3} and we obtain a equation as follows:
\begin{align}
	0=&
%	-2\mu^2 v_1+4\lambda_1 v_1(v_1^2+2v_2v_3)+4\lambda_2 v_3(v_2^2+2v_3v_1)+4\lambda_3 v_2(v_3^2+2v_1v_2) \notag \\
%	&+4\lambda_4\Big[v_1(v_1^2-v_2v_3)-\frac{1}{2}v_2(v_3^2-v_1v_2)-\frac{1}{2}v_3(v_2^2-v_3v_1)\Big] \notag \\
%	&-2\mu^2 v_2+4\lambda_1 v_3(v_1^2+2v_2v_3)+4\lambda_2 v_2(v_2^2+2v_3v_1)+4\lambda_3 v_1(v_3^2+2v_1v_2) \notag \\
%	&+4\lambda_4\Big[v_2(v_2^2-v_3v_1)-\frac{1}{2}v_1(v_3^2-v_1v_2)-\frac{1}{2}v_3(v_1^2-v_2v_3)\Big] \notag \\
%	&-2\mu^2 v_3+4\lambda_1 v_2(v_1^2+2v_2v_3)+4\lambda_2 v_1(v_2^2+2v_3v_1)+4\lambda_3 v_3(v_3^2+2v_1v_2) \notag \\
%	&+4\lambda_4\Big[v_3(v_3^2-v_1v_2)-\frac{1}{2}v_1(v_2^2-v_3v_1)-\frac{1}{2}v_2(v_1^2-v_2v_3)\Big] \notag \\
	(v_1+v_2+v_3) \Big[ -2\mu^2 + 4\lambda_1 (v_1^2 + 2v_1v_2) + 4\lambda_2 (v_2^2 + 2v_3v_1) + 4\lambda_3 (v_3^2 + 2v_1v_2) \notag \\
	&+4\lambda_4(v_1^2 + v_2^2 + v_3^2 - v_1v_2 - v_2v_3 - v_3v_1) \Big] \notag \\
	=&4 (v_1+v_2+v_3) \Big[ (\lambda_1+\lambda_4)v_1^2 + \big\{(2\lambda_3-\lambda_4)v_2 + (2\lambda_2-\lambda_4)v_3 \big\} v_1 \notag \\ 
	&+ (\lambda_2+\lambda_4)v_2^2 + (\lambda_3+\lambda_4)v_3^2 + (2\lambda_1-\lambda_4)v_2v_3 -\frac{\mu^2}{2} \Big].
\end{align}
When $v_1+v_2+v_3 \ne 0$ is satisfied, we obtain 
\begin{align} \label{eq:vacuum_eq}
	&(\lambda_1+\lambda_4)v_1^2 + \big\{(2\lambda_3-\lambda_4)v_2 + (2\lambda_2-\lambda_4)v_3 \big\} v_1 \notag \\
	&+ (\lambda_2+\lambda_4)v_2^2 + (\lambda_3+\lambda_4)v_3^2 + (2\lambda_1-\lambda_4)v_2v_3 -\frac{\mu^2}{2}=0.
\end{align}
If $\lambda_1+\lambda_4$ which is the coefficient of $v_1^2$ in Eq.~\eqref{eq:vacuum_eq} holds zero, we find the solution $(1,1,1)$.
In this case we cannot realize the current experimental data.
When $\lambda_1+\lambda_4\ne0$, $\lambda_3=\lambda_2$, and $2\lambda_1+2\lambda_2+\lambda_4\neq0$ hold, 
we obtain the following solutions:
\begin{align}
\label{eq:vev1}
	\langle\phi_1\rangle=&v_1 ,\\
\label{eq:vev2}
	\langle\phi_2\rangle=&v_2 \\
  =&-\frac{2\lambda_2-\lambda_4}{2\lambda_1+2\lambda_2+\lambda_4}v_1\nonumber \\
	&\pm\frac{\sqrt{\{-2\lambda_1^2+2\lambda_2^2-2\lambda_2(\lambda_1-3\lambda_4)-3\lambda_1\lambda_4\}v_1^2+\frac{1}{2}\left (2\lambda_1+2\lambda_2+\lambda_4\right )\mu^2}}{2\lambda_1+2\lambda_2+\lambda_4},\nonumber \\
\label{eq:vev3}
  \langle\phi_3\rangle=&v_3 \\
  =&-\frac{2\lambda_2-\lambda_4}{2\lambda_1+2\lambda_2+\lambda_4}v_1\nonumber \\
	&\pm\frac{\sqrt{\{-2\lambda_1^2+2\lambda_2^2-2\lambda_2(\lambda_1-3\lambda_4)-3\lambda_1\lambda_4\}v_1^2+\frac{1}{2}\left (2\lambda_1+2\lambda_2+\lambda_4\right )\mu^2}}{2\lambda_1+2\lambda_2+\lambda_4}. \nonumber 
\end{align}
These VEVs can be written as
\begin{equation}
\label{eq:VEVs_A4}
v_1=v\cos{\beta},  \quad v_2=\frac{v}{\sqrt2}\sin{\beta},  \quad v_3=\frac{v}{\sqrt2}\sin{\beta},
\end{equation}
where, the range of $\lambda$ are $\lambda_1<\lambda_2$ and $\lambda_1<\lambda_4$.
These $\lambda$ conditions are derived from minimum conditions of Higgs potential.

%In Refs.~\cite{Degee:2012sk,GonzalezFelipe:2013yhh} the 3HDM potential analysis with 
%$A_4$ symmetry are also discussed. Since their bases are different compared with us 
%in Eqs.~\eqref{eq:one-dimensional_rep} and~\eqref{eq:three-dimensional_rep}, 
%our forms of the VEVs for the potential analysis Eqs.~\eqref{eq:vev1}-\eqref{eq:VEVs_A4} 
%are different compared with theirs~\cite{Degee:2012sk,GonzalezFelipe:2013yhh}. 
%In the point of the view for the phenomenological aspects, we can discuss the independent on the bases. 
Note that there are other two solution forms in Eq.~\eqref{eq:VEVs_A4}.
When $\lambda_3=\lambda_1$, $\lambda_2+\lambda_4\ne0$ and 
$2\lambda_1+2\lambda_2+\lambda_4\neq0$ hold, we obtain 
\begin{equation}
\label{eq:VEVs_A4_2}
v_1=\frac{v}{\sqrt2}\sin{\beta},  \quad v_2=v\cos{\beta},  \quad v_3=\frac{v}{\sqrt2}\sin{\beta}.
\end{equation}
On the other hand, when $\lambda_2=\lambda_1$, $\lambda_3+\lambda_4\ne0$ and 
$2\lambda_1+2\lambda_3+\lambda_4\neq0$ hold, we obtain 
\begin{equation}
\label{eq:VEVs_A4_3}
v_1=\frac{v}{\sqrt2}\sin{\beta},  \quad v_2=\frac{v}{\sqrt2}\sin{\beta},  \quad v_3=v\cos{\beta}.
\end{equation}
In the next section, we present our $A_4$ model and caluculate the mass matries.

%------------------------------------------------------------------------------%
%------------------------------------------------------------------------------%
%------------------------------------------------------------------------------%
\section{Lepton flavor model in the $A_4$ symmetry}
\label{sec:model}
In this section, we present a non-SUSY lepton flavor model in the $A_4$ symmetry. 
The left-handed lepton doublets are assigned to triplet and the right-handed 
charged leptons are assigned to different singlets as ${\bf 1}$, ${\bf 1''}$, and ${\bf 1'}$ 
of the $A_4$ symmetry, respectively.
We introduce the right-handed Majorana neutrinos which are assigned to triplet 
of the $A_4$ symmetry. We also introduce three Higgs doublets which are assigned as triplet 
of the $A_4$ symmetry as discussed in Section~\ref{sec:3HDM}.
In Table~\ref{tab:model}, we summarize the particle assignments 
of $SU(2)_L$ and $A_4$ symmetry\footnote{In the AF 
model~\cite{Altarelli:2005yp,Altarelli:2005yx}, they introduce 
the $Z_{3}$ symmetry in order to obtain the relevant couplings. 
Thanks for the $SU(2)_L$ gauge symmetry, we do not need to add the extra symmetry 
in our model.}.
\begin{table}[h]
  \centering
  \begin{tabular}{|c||ccccc|c|}
    \hline 
    \rule[14pt]{0pt}{0pt}
    & $\bar{\ell}=(\bar{\ell}_e,\bar{\ell}_\mu,\bar{\ell}_\tau)$ & $e_R$ &$\mu_R$ & $\tau_R$ & 
    $\nu_R=(\nu_{R1},\nu_{R2},\nu_{R3})$ & $\Phi=(\phi_1,\phi_2,\phi_3)$
    \\ \hline 
    \rule[14pt]{0pt}{0pt}
    $SU(2)_L$ & $2$ & $1$ & $1$ & $1$ & $1$ & 2
    \\
    $A_4$ & ${\bf 3}$ & ${\bf 1}$ & ${\bf 1''}$ & ${\bf 1'}$ & ${\bf 3}$ & ${\bf 3}$\\
    %$Z_3$ & $1$ & $1$ & $\omega^2$ & $\omega$ & $1$ & $\omega$ \\
    \hline
  \end{tabular}
  \caption{The charge assignments of $SU(2)_L\times A_4$ symmetry in our model.}
%, $\ell$ is 3 lepton doublets.}
  \label{tab:model}
\end{table}

We can write down the Lagrangian for Yukawa interactions and Majorana mass term 
in our model. The $SU(2)_L\times A_4$ invariant Lagrangian is written as
%Yukawa interaction and Majorana mass term in Lagrangian for this model is Table~\ref{tab:model}
\begin{equation}\label{eq:lagrangina_all}
  \mathcal{L}_Y = \mathcal{L}_{\ell} + \mathcal{L}_D + \mathcal{L}_M + h.c. ,
\end{equation}
where 
\begin{align}
\label{eq:lagrangian_lepton}
\mathcal{L}_{\ell} &= y_e\bar{\ell}\Phi e_R + y_\mu\bar{\ell}\Phi \mu_R 
+ y_\tau\bar{\ell}\Phi \tau_R, \\
\label{eq:lagrangian_dirac}
\mathcal{L}_D &= y_D\bar{\ell}\tilde{\Phi}\nu_R, \\
\label{eq:lagrangian_majorana}
\mathcal{L}_M &= \frac{1}{2} M\bar{\nu}_R^c\nu_R.
\end{align}
%\begin{equation}\label{eq:lagrangian_lepton}
%  \mathcal{L}_{\ell} = y_e\bar{\ell}\phi e_R + y_\mu\bar{\ell}\phi \mu_R + y_\tau\bar{\ell}\phi \tau_R,
%\end{equation}
%\begin{equation}\label{eq:lagrangian_dirac}
%  \mathcal{L}_D = y_D\bar{\ell}\tilde{\phi}\nu_R,
%\end{equation}
%\begin{equation}\label{eq:lagrangian_majorana}
%  \mathcal{L}_M = \frac{1}{2} M\bar{\nu}_R^c\nu_R.
%\end{equation}
Note that $y_e , y_\mu , y_\tau$, and $y_D$ are Yukawa couplings and 
$M$ is the right-handed Majorana neutrino mass.
%$\phi=(\phi_1,\phi_2,\phi_3)$ are Higgs doublets. 
After the SSB, three Higgs doublets have VEVs as
$\braket{\Phi}=(v_1,v_2,v_3)$. 
In the charged lepton sector Eq.~\eqref{eq:lagrangian_lepton}, 
the Yukawa interactions are rewritten as
%　 After spontaneous symmetry breaking, 3 Higgs doublet have vaccum expectation value $\braket{\phi}=\left(\begin{pmatrix}0\\v1\end{pmatrix},\begin{pmatrix}0\\v2\end{pmatrix},\begin{pmatrix}0\\v3\end{pmatrix}\right)$. First we consider charaged lepton mass matrix. Eq.\eqref{eq:lagrangian_lepton} have following form,
\begin{align}
  y_e\bar{\ell}\Phi e_R &= y_e(\bar{\ell}_e\phi_1+\bar{\ell}_\mu\phi_3+\bar{l}_\tau\phi_2)e_R
  \\&=
  y_e(\bar e_L v_1 + \bar \mu_L v_3 + \bar \tau_L v_2)e_R,\nonumber \\
  y_\mu\bar{\ell}\Phi\mu_R &= y_\mu(\bar{\ell}_\tau\phi_3 + \bar{\ell}_e\phi_2 + \bar{\ell}_\mu\phi_1)\mu_R
  \\&=
 y_\mu(\bar \tau_L v_3+\bar e_L v_2 + \bar \mu_L v_1)\mu_R,\nonumber \\
  y_\tau\bar{\ell}\Phi\tau_R &= y_\tau(\bar{\ell}_\mu\phi_2 + \bar{\ell}_e\phi_3 + \bar{\ell}_\tau\phi_1)\tau_R
  \\&=
 y_\tau(\bar \mu_L v_2+\bar e_L v_3 + \bar \tau_L v_1)\tau_R.\nonumber
\end{align}
Then, the charged lepton mass matrix $M_{\ell }$ is obtained as
\begin{equation}\label{eq:mass_matrix_charged_lepton}
  M_{\ell}=
  \begin{pmatrix}
    y_e v_1 & y_\mu v_2 & y_\tau v_3 \\
    y_e v_3 & y_\mu v_1 & y_\tau v_2 \\
    y_e v_2 & y_\mu v_3 & y_\tau v_1 \\
  \end{pmatrix}_{LR}.
\end{equation}
Here and hereafter we take the left-right basis in the mass matrices. 
In order to obtain the left-side unitary mixing matrix $U_\ell $, we consider $M_\ell M_\ell ^\dagger $ as
\begin{align}
\label{eq:MldaggerMl}
&M_\ell M_\ell ^\dagger = \\
&\begin{pmatrix}
|y_e|^2v_1^2+|y_\mu |^2v_2^2+|y_\tau |^2v_3^2 & 
|y_e|^2v_1v_3+|y_\mu |^2v_1v_2+|y_\tau |^2v_2v_3 &
|y_e|^2v_1v_2+|y_\mu |^2v_2v_3+|y_\tau |^2v_1v_3 \\
%|y_e|^2v_1v_3+|y_\mu |^2v_1v_2+|y_\tau |^2v_2v_3 & 
\dots & |y_e|^2v_3^2+|y_\mu |^2v_1^2+|y_\tau |^2v_2^2 &
|y_e|^2v_2v_3+|y_\mu |^2v_1v_3+|y_\tau |^2v_1v_2 \\
%|y_e|^2v_1v_2+|y_\mu |^2v_2v_3+|y_\tau |^2v_1v_3 & 
%|y_e|^2v_2v_3+|y_\mu |^2v_1v_3+|y_\tau |^2v_1v_2 & 
\dots & \dots & |y_e|^2v_2^2+|y_\mu |^2v_3^2+|y_\tau |^2v_1^2
\end{pmatrix}. \nonumber 
\end{align}
Then, the elements of the mass matrix Eq.~\eqref{eq:MldaggerMl} are real 
and we can numerically calculate the charged lepton masses in Section~\ref{sec:Numerical}. 
Next, we derive the Dirac neutrino mass matrix from Eq.~\eqref{eq:lagrangian_dirac}. 
By using the $A_4$ multiplication rule (See appendix~\ref{sec:multiplication-rule}.), 
we can make the singlet term. 
When we take the complex conjugate, we need to take care 
$\tilde{\Phi}=(\tilde{\phi}_1,\tilde{\phi}_3,\tilde{\phi}_2)$ because of the complex conjugate for 
the generator $T$ in Eq.~\eqref{eq:complex_generator_T}. 
Since Eq.~\eqref{eq:lagrangian_dirac} is $3\otimes3\otimes3$ form in $A_4$ symmetry, 
we first make $3_{S}\oplus3_{A}\in3\otimes3$ as
\begin{equation}
  y_D\bar{\ell}\tilde{\Phi} =
  \frac{y_{DS}}{3}
  \begin{pmatrix}
    2\bar{\ell}_e \tilde{\phi}_1 - \bar{\ell}_\mu \tilde{\phi}_2 - \bar{\ell}_\tau \tilde{\phi}_3 \\
    2\bar{\ell}_\tau \tilde{\phi}_2 - \bar{\ell}_e \tilde{\phi}_3 - \bar{\ell}_\mu \tilde{\phi}_1 \\
    2\bar{\ell}_\mu \tilde{\phi}_3 - \bar{\ell}_\tau \tilde{\phi}_1 - \bar{\ell}_e \tilde{\phi}_2
  \end{pmatrix}_{3_S}
  +\frac{y_{DA}}{2}
  \begin{pmatrix}
    \bar{\ell}_\mu \tilde{\phi}_2 - \bar{\ell}_\tau \tilde{\phi}_3 \\
    \bar{\ell}_e \tilde{\phi}_3 - \bar{\ell}_\mu \tilde{\phi}_1 \\
    \bar{\ell}_\tau \tilde{\phi}_1 - \bar{\ell}_e \tilde{\phi}_2 \\
  \end{pmatrix}_{3_A},
\end{equation}
where $y_{DS}$ and $y_{DA}$ are the symmetric and anti-symmetric Dirac Yukawa couplings, 
respectively. 
Then, the Dirac neutrino Yukawa interaction in Eq.~\eqref{eq:lagrangian_dirac} 
can be written as follows:
\begin{align}
  y_D\bar{\ell}\tilde{\Phi}\nu_R
  =&
    \frac{y_{DS}}{3}
    \begin{pmatrix}
      2\bar{\ell}_e \tilde{\phi}_1 - \bar{\ell}_\mu \tilde{\phi}_2 - \bar{\ell}_\tau \tilde{\phi}_3 \\
      2\bar{\ell}_\tau \tilde{\phi}_2 - \bar{\ell}_e \tilde{\phi}_3 - \bar{\ell}_\mu \tilde{\phi}_1 \\
      2\bar{\ell}_\mu \tilde{\phi}_3 - \bar{\ell}_\tau \tilde{\phi}_1 - \bar{\ell}_e \tilde{\phi}_2
    \end{pmatrix}_{3_S}\otimes 
    \begin{pmatrix}
      \nu_{R1} \\ \nu_{R2} \\ \nu_{R3}
    \end{pmatrix}
    +\frac{y_{DA}}{2}
    \begin{pmatrix}
      \bar{\ell}_\mu \tilde{\phi}_2 - \bar{\ell}_\tau \tilde{\phi}_3 \\
      \bar{\ell}_e \tilde{\phi}_3 - \bar{\ell}_\mu \tilde{\phi}_1 \\
      \bar{\ell}_\tau \tilde{\phi}_1 - \bar{\ell}_e \tilde{\phi}_2 \\
    \end{pmatrix}_{3_A}\otimes 
    \begin{pmatrix}
      \nu_{R1} \\ \nu_{R2} \\ \nu_{R3}
    \end{pmatrix}
  \\=&
    \frac{y_{DS}}{3}\left[
    (2\bar{\ell}_e \tilde{\phi}_1 - \bar{\ell}_\mu \tilde{\phi}_2 - \bar{\ell}_\tau \tilde{\phi}_3)\nu_{R1}
    + (2\bar{\ell}_\tau \tilde{\phi}_2 - \bar{\ell}_e \tilde{\phi}_3 - \bar{\ell}_\mu \tilde{\phi}_1)\nu_{R3}
    \right.
    \\&\qquad \qquad
    + \left.(2\bar{\ell}_\mu \tilde{\phi}_3 - \bar{\ell}_\tau \tilde{\phi}_1 - \bar{\ell}_e \tilde{\phi}_2)\nu_{R2} \right]
  \nonumber \\ &
  +\frac{y_{DA}}{2} \left[ (\bar{\ell}_\mu \tilde{\phi}_2 - \bar{\ell}_\tau \tilde{\phi}_3)\nu_{R1} + (\bar{\ell}_e \tilde{\phi}_3 - \bar{\ell}_\mu \tilde{\phi}_1)\nu_{R3} + (\bar{\ell}_\tau \tilde{\phi}_1 - \bar{\ell}_e \tilde{\phi}_2)\nu_{R2} \right],\nonumber
\end{align}
where $\tilde{\Phi}= -i\sigma_2\Phi ^*$ and the  VEVs of 
$\tilde{\phi}_i$ are $\braket{\tilde{\phi _i}}=v_i$.
Therefore, the Dirac neutrino mass matrix $M_D$ is obtained as
\begin{equation}
\label{eq:mass_matrix_dirac_neutrino}
  M_D =
  \frac{y_{DS}}{3}
  \begin{pmatrix}
    2v_1 & -v_2 & -v_3 \\
    -v_2 & 2v_3 & -v_1 \\
    -v_3 & -v_1 & 2v_2
  \end{pmatrix}_{LR}
  + \frac{y_{DA}}{2}
  \begin{pmatrix}
    0 & -v_2 & v_3 \\
    v_2 & 0 & -v_1 \\
    -v_3 & v_1 & 0
  \end{pmatrix}_{LR}.
\end{equation}
Next we discuss the right-handed Majorana neutrino mass matrix. 
In Eq.~\eqref{eq:lagrangian_majorana}, the right-handed Majorana neutrino mass term is 
decomposed as follows:
\begin{equation}
M\bar{\nu}_R^c \nu_R=M(\bar \nu_{R1}^c\nu_{R1} + \bar \nu_{R2}^c\nu_{R3} + 
\bar \nu_{R3}^c\nu_{R2}).
\end{equation}
Then, the right-handed Majorana neutrino mass matrix $M_R$ is
\begin{equation}
\label{eq:mass_matrix_majorana_neutrino}
  M_R =M
  \begin{pmatrix}
    1 & 0 & 0 \\
    0 & 0 & 1 \\
    0 & 1 & 0
  \end{pmatrix}.
\end{equation}
%Therefore, we obtain the neutrino mass matrices for the Dirac 
%Eq.~\eqref{eq:mass_matrix_dirac_neutrino} and Majorana 
%Eq.~\eqref{eq:mass_matrix_majorana_neutrino} ones,
%\begin{equation}\label{eq:neutrino_all_mass_matrix}
%  \begin{matrix}
%     & \begin{matrix} L && R\end{matrix} \\
%     \begin{matrix}L \\ R\end{matrix} &
%       \begin{pmatrix} 0 & M_D \\ M_D^T & M_R\end{pmatrix}.
%  \end{matrix}
%\end{equation}
%Neutrino block mass matrix Eq~\eqref{eq:neutrino_all_mass_matrix} can be easily diagonarized, and block diagonarized components is left-handed neutrino mass matrix.
By using the type-I seesaw mechanism~\cite{Minkowski:1977sc,Yanagida,Gell-Mann:1979vob,Mohapatra:1979ia,Schechter:1980gr}, the left-handed Majorana neutrino mass matrix 
$m_\nu$ is written as 
\begin{equation}\label{eq:mass_matrix_left_neutrino}
  m_\nu=-M_DM_R^{-1}M_D^T
\end{equation}
\begin{align} 
  =&
  \frac{1}{M}
  \begin{pmatrix}
    2v_2v_3(y_{DA}^2-y_{DS}^2) & v_1v_2(y_{DS}^2-4y_{DS}y_{DA}-y_{DA}^2) & v_1v_3(y_{DS}^2+4y_{DS}y_{DA}-y_{DA}^2)
    \\
    v_1v_2(y_{DS}^2-4y_{DS}y_{DA}-y_{DA}^2) & 4v_1v_3y_{DS}(y_{DS}+y_{DA}) & -y_{DS}^2(v_1^2+5v_2v_3)
    \\
    v_1v_3(y_{DS}^2+4y_{DS}y_{DA}-y_{DA}^2) & -y_{DS}^2(v_1^2+5v_2v_3) & 4v_1v_2y_{DS}(y_{DS}-y_{DA})
  \end{pmatrix}
 \nonumber  \\&\quad +
  \frac{1}{M}
  \begin{pmatrix}
    -4 v_1^2 y_{DS}^2 & 2v_3^2y_{DS}(y_{DS}-y_{DA}) & 2v_2^2y_{DS}(y_{DS}+y_{DA})
    \\
    2v_3^2y_{DS}(y_{DS}-y_{DA}) & -v_2^2(y_{DA}-y_{DS})^2 & y_{DA}^2(v_1^2+v_2v_3)
    \\
    2v_2^2y_{DS}(y_{DS}+y_{DA}) & y_{DA}^2(v_1^2+v_2v_3) & -v_3^2(y_{DA}-y_{DS})^2
  \end{pmatrix}, \notag
\end{align}
where we redefine the Dirac Yukawa couplings 
$y_{DS}/3\to y_{DS}$ and $y_{DA}/2\to y_{DA}$ for simplicity.
In our model, the Dirac neutrino mass matrix has symmetric and anti-symmetric 
Yukawa couplings for the $A_4$ symmetry in Eq.~\eqref{eq:mass_matrix_dirac_neutrino}.
On the other hand, in Eq.~\eqref{eq:mass_matrix_majorana_neutrino}, 
the right-handed Majorana neutrino mass matrix has a simple flavor structure, 
c.f., in the AF Model the Dirac neutrino mass matrix is simple. On the other hand, 
the right-handed Majorana neutrino mass matrix is the structure which derives the TBM 
in their model. In the next section, we show the numerical analysis.

%------------------------------------------------------------------------------%
\section{Numerical analysis}% 20220910 Takei added
\label{sec:Numerical}
%------------------------------------------------------------------------------%
In this section, we show the numerical analysis such as the lepton flavor mixing angles, 
Dirac CP phase, Majorana phases, and the effective mass for the $0\nu \beta \beta $ decay. 
We discuss what is our model verifiable in the near future experiments.
%In this section, we predict lepton flavor mixing angles, Dirac CP phase and Majorana phases. We discuss what is this model verifable in the future.

\begin{figure}[t]
\begin{minipage}{0.5\hsize}
  \begin{center}
  \includegraphics[width=75mm]{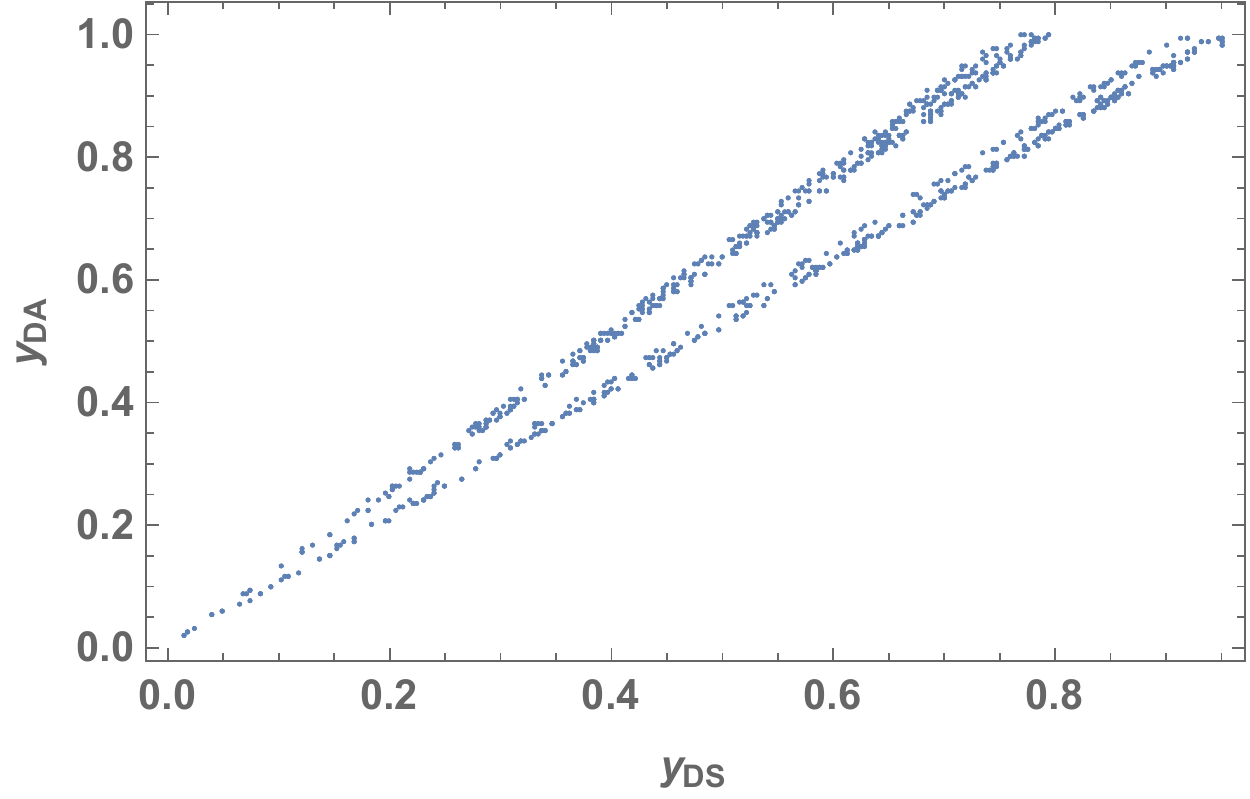}
  \subcaption{}
  \label{fig:neutrino_yukawa}
  \end{center}
\end{minipage}
\begin{minipage}{0.5\hsize}
  \centering
  \includegraphics[width=75mm]{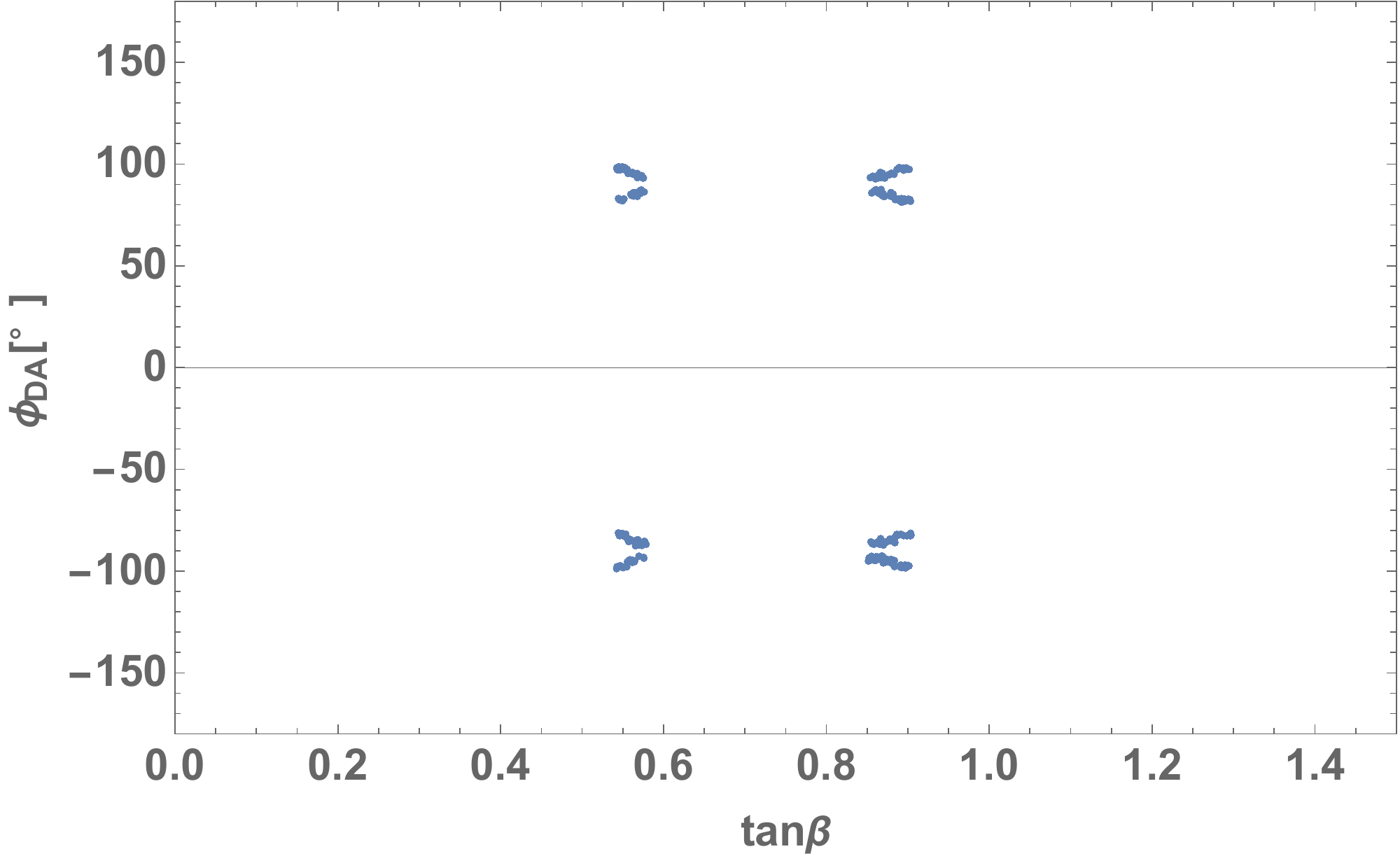}
  \subcaption{}
  \label{fig:tan_yukawa}%Relation between the phase of $y_{DA}$ and Higgs VEV.}
\end{minipage}
\caption{(a)~The relation between the Dirac neutrino Yukawa couplings 
$y_{DS}$ and $y_{DA}$.
We simulate by assigning different numerical values to the Dirac neutrino Yukawa couplings 
at $\mathcal{O}(1)$.
(b)~Higgs VEV ratio $\tan \beta$ and complex phase of the Dirac Yukawa coupling 
$\phi_{DA}$.}
\end{figure}

In section~\ref{sec:3HDM}, we have analyzed the Higgs potential 
and found the following solution\footnote{
The solution form in Eq.~\eqref{eq:VEVs_A4_2} can be realised by taking charge assignments such that 
the left-handed lepton doublets are assigned to triplet $\bar{\ell}=(\bar{\ell}_\tau,\bar{\ell}_\mu,\bar{\ell}_e)$ and the right-handed 
charged leptons are assigned to different singlets as ${\bf 1}$, ${\bf 1'}$, and ${\bf 1''}$ and the right-handed Majorana neutrinos are assigned to triplet as $\nu_R=(\nu_{R3},\nu_{R2},\nu_{R1})$.
On the other hand, the solution form in Eq.~\eqref{eq:VEVs_A4_3} can be realised by taking charge assignments such that 
the left-handed lepton doublets are assigned to triplet $\bar{\ell}=(\bar{\ell}_\mu,\bar{\ell}_e,\bar{\ell}_\tau)$ and the right-handed 
charged leptons are assigned to different singlets as ${\bf 1}$, ${\bf 1'}$, and ${\bf 1''}$ and the right-handed Majorana neutrinos are assigned to triplet as $\nu_R=(\nu_{R2},\nu_{R1},\nu_{R3})$. In these solution forms we obtain same numerical results in section~\ref{sec:Numerical}.}
which was derived from Higgs potential minimization as 
\begin{equation}\label{eq:higgs_vev}
  (v_1,v_2,v_3) = ( v\cos\beta ,\frac{v}{\sqrt{2}}\sin\beta ,\frac{v}{\sqrt{2}}\sin\beta ).
\end{equation}
Since charged lepton mass matrix Eq.~\eqref{eq:MldaggerMl} is only 
depend on three Yukawa couplings and Higgs VEVs.
Then once we fix the Higgs VEVs, we can obtain charged lepton Yukawa couplings 
by solving the following equations and the unitary matrix which diagonalizes charged lepton 
mass matrix.
\begin{align}\label{eq:mass_equation}
    \mathrm{Tr}(M_\ell M_\ell^\dag) &= m_e^2 + m_\mu^2 + m_\tau^2, \nonumber \\
    \mathrm{Det}(M_\ell M_\ell^\dag) &= m_e^2 m_\mu^2 m_\tau^2, \\
    \left(\mathrm{Tr}M_\ell M_\ell^\dag\right)^2 -\mathrm{Tr}\left [(M_\ell M_\ell^\dag)^2\right ]
   &=2(m_e^2m_\mu^2 + m_\mu^2m_\tau^2 + m_\tau^2m_e^2), \nonumber 
\end{align}
where $m_e, m_\mu , m_\tau$ are charged lepton masses.
The Dirac neutrino Yukawa couplings similarly obtained by solving equations which are 
substituted Eq.~\eqref{eq:mass_matrix_left_neutrino} into  Eq.~\eqref{eq:mass_equation}, 
where we use the left-handed Majorana neutrino masses instead of the charged lepton masses 
in the right-side of Eq.~\eqref{eq:mass_equation}.
However neutrino masses are only known mass squared differences 
$\Delta m_{21}^2$ and $\Delta m_{32(31)}^2$ in the inverted (normal) ordering of the neutrino 
mass hierarchy. 
Then we need to decide the Dirac neutrino Yukawa couplings $y_{DS}$ and $y_{DA}$ 
to satisfy experimental data in Table~\ref{table:NuFIT_data}.
%\footnote{We show only 
%iverted ordering case because we found the only inverted ordering case 
%in our numerical calculation.}. 
%-------------NuFIT 5.1 data--------------
\begin{table}[t]
  \centering
  \begin{tabular}{|c|cc|}
  \hline 
  \rule[14pt]{0pt}{0pt}
    Inverted Ordering & bfp $\pm1\sigma$ & $3\sigma$ range \\
     \hline
     \rule[14pt]{0pt}{0pt}
    $\sin^2\theta_{12}$ & $0.304^{+0.013}_{-0.012}$ & $0.269 \to 0.343$ \\ \rule[14pt]{0pt}{0pt}
    $\theta_{12}/^\circ$ & $33.45^{+0.78}_{-0.75}$ & $31.27\to35.87$ \\ \rule[14pt]{0pt}{0pt}
    $\sin^2\theta_{23}$ & $0.570^{+0.016}_{-0.022}$ & $0.410\to0.613$ \\ \rule[14pt]{0pt}{0pt}
    $\theta_{23}/^\circ$ & $49.0^{+0.9}_{-1.3}$ & $39.8\to51.6$ \\ \rule[14pt]{0pt}{0pt}
    $\sin^2\theta_{13}$ & $0.02241^{+0.00074}_{-0.00062}$ & $0.02055\to0.02457$ \\ \rule[14pt]{0pt}{0pt}
    $\theta_{13}/^\circ$ & $8.61^{+0.14}_{-0.12}$ & $8.24\to9.02$ \\ \rule[14pt]{0pt}{0pt}
    $\delta_{\mathrm{CP}}/^\circ$ & $278^{+22}_{-30}$ & $194\to345$ \\ \rule[14pt]{0pt}{0pt}
    $\frac{\Delta m_{21}^2}{10^{-5} [\mathrm{eV}^2]}$ & $7.42^{+0.21}_{-0.20}$ & $6.82\to8.04$\\ \rule[16pt]{0pt}{0pt}
    $\frac{\Delta m_{32}^2}{10^{-3} [\mathrm{eV}^2]}$ & $-2.490^{+0.026}_{-0.028}$ & $-2.574\to-2.410$ \rule[-8pt]{0pt}{0pt} \\ \hline 
  \end{tabular}
  \caption{NuFIT 5.1 data~\cite{Esteban:2020cvm,Gonzalez-Garcia:2021dve} 
  in the inverted neutirno mass ordering, 
  where $\Delta m^2_{ij}$ is mass squared difference between $m_{i}$ and $m_{j}$. }
  \label{table:NuFIT_data}
\end{table}
In our numerical analysis, we simulate by assigning different numerical values to 
the lightest neutrino mass at the range $m_3\in[0,15.9]$~[meV], 
where the upper limit of the lightest neutrino mass 
is estimated in Ref.~\cite{Planck:2018vyg}.
%Here we assume the Majorana mass is $\mathcal{O}(10^{15})$~GeV.
We analyze our model in normal and inverted neutrino mass orderings.
When we assume the normal ordering, there are no realistic parameters which satisfy 
the current experimental data in Table~\ref{table:NuFIT_data}. 
We then assume the inverted ordering for the rest of our discussion.
In Fig.~\ref{fig:neutrino_yukawa}, we show the symmetric and anti-symmetric Dirac 
Yukawa couplings for Eq.~\eqref{eq:mass_matrix_dirac_neutrino} which 
are satisfied the current experimental data in Table~\ref{table:NuFIT_data}. 
These Yukawa couplings look like proportional each other and $\mathcal{O}(1)$. 
The Majorana mass $M$ in Eq.~\eqref{eq:mass_matrix_majorana_neutrino} is 
allowed at $\mathcal{O}(10^{15})$~[GeV].
We also show the numerical results of the Higgs VEV ratio $\tan \beta$ and 
the complex phase of the Dirac Yukawa coupling in Fig.~\ref{fig:tan_yukawa}. 
The $\tan \beta$ is localized around $\tan\beta\in[0.4,0.6],[0.8,1.2]$.
The complex phase of the Dirac Yukawa coupling $\phi_{DA}$ only appear in 
$70[{\ }^\circ]$~-~$100[{\ }^\circ]$. 
In our model, the complex phase of the Dirac Yukawa coupling which contribute to the 
mixing matrix is only $\phi_{DA}$, then this result has a strong influence to the CP phases.
%We analyze these CP phases below.\

\begin{figure}[h]
    \begin{center}
    \includegraphics[width=75mm]{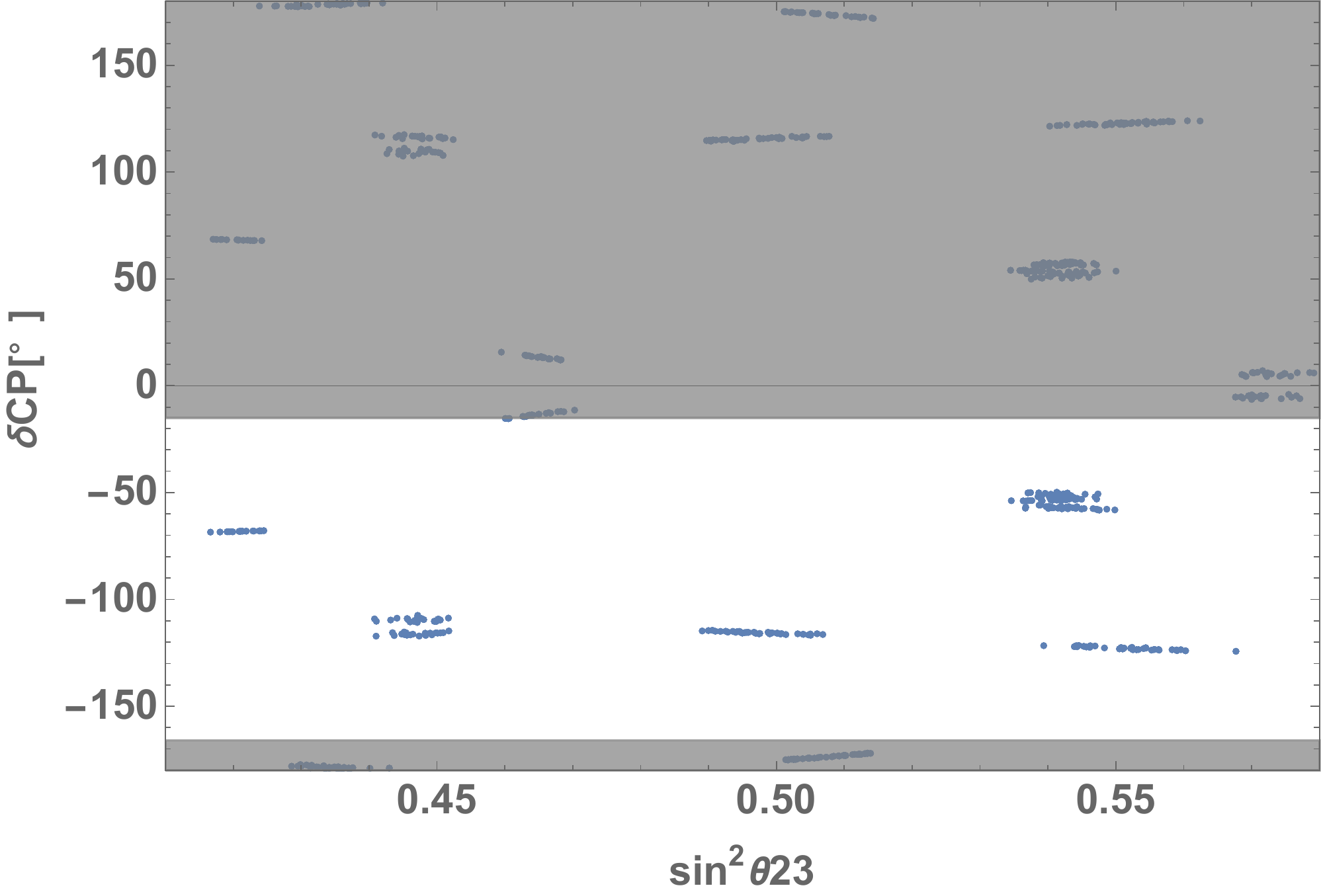}
    \caption{The allowed region for the $\sin ^2\theta _{23}$ and $\delta _\text{CP}$ 
    within $3\sigma $ standard deviation of the Table~\ref{table:NuFIT_data}.
    The gray area is outside of  $3\sigma$ standard deviation in %global analysis 
    NuFIT 5.1 data~\cite{Esteban:2020cvm,Gonzalez-Garcia:2021dve}.}
    \label{fig:23_cp}
    \end{center}
\end{figure}

%From above analysis we obtain mass matrices.
In the PDG parametrization, we can write the PMNS matrix as 
%In our calculation, we can obtain two unitary matrices $U_{\ell}, U_{\nu}$ and a Majorana phase matrix $P_\nu$. 
%which diagonalize mass matrices Eq.~\eqref{eq:mass_matrix_charged_lepton} and 
%Eq.~\eqref{eq:mass_matrix_left_neutrino} and also obtain the PMNS matrix 
%$U_{\mathrm{PMNS}}= U_{\ell}^\dag U_{\nu}P_\nu$,
\begin{align}
\label{eq:PMNS_matrix}
  &\qquad \qquad U_{\mathrm{PMNS}}^{\text{PDG}} =
  \begin{pmatrix}
    U_{e1} & U_{e2} & U_{e3} \\
    U_{\mu1} & U_{\mu2} & U_{\mu3} \\
    U_{\tau1} & U_{\tau2} & U_{\tau3}
  \end{pmatrix}
  \\&=
  \begin{pmatrix}
    c_{12}c_{13} & s_{12}c_{13} & s_{13}\mathrm{e}^{-i\delta_{\mathrm{CP}}} \\
    -s_{12}c_{23}-c_{12}s_{23}s_{13}\mathrm{e}^{i\delta_{\mathrm{CP}}} & c_{12}c_{23}-s_{12}s_{23}s_{13}\mathrm{e}^{i\delta_{\mathrm{CP}}} & s_{23}c_{13} \\
    s_{12}s_{23}-c_{12}c_{23}s_{13}\mathrm{e}^{i\delta_{\mathrm{CP}}} & -c_{12}s_{23}-s_{12}c_{23}s_{13}\mathrm{e}^{i\delta_{\mathrm{CP}}} & c_{23}c_{13}
  \end{pmatrix}
  \begin{pmatrix}
    \mathrm{e}^{i\eta_1}&&\\
    &\mathrm{e}^{i\eta_2}&\\
    &&1 \\
  \end{pmatrix}, \nonumber 
\end{align}
where $U_{\alpha i}~(\alpha =e,\mu , \tau ,~i=1,2,3)$ are the PMNS matrix elements, 
$c_{ij}$ and $s_{ij}$ denote $\cos \theta _{ij}$ and $\sin \theta _{ij}$, $\delta _\text{CP}$ is the Dirac CP phase, and $\eta_1$ and $\eta_2$ are Majorana phases, respectively~\cite{Workman:2022ynf}. 
The lepton mixing angles are obtained as follows:
\begin{equation}
\sin \theta _{13}=\left |U_{e3}\right |,\quad 
\tan \theta _{12}=\left |\frac{U_{e2}}{U_{e1}}\right |,\quad 
\tan \theta _{23}=\left |\frac{U_{\mu 3}}{U_{\tau 3}}\right |.
\end{equation}
In addition, the Dirac CP phase is determined by one of the Jarlskog Invariants~\cite{Jarlskog:1985ht} as
\begin{equation}
J_\text{CP}=\text{Im}\left [U_{e1}U_{e2}^*U_{\mu 2}U_{\mu 1}^*\right ],
\end{equation}
and 
\begin{equation}
J_\text{CP}=\sin \theta _{12}\cos \theta _{12}\sin \theta _{23}\cos \theta _{23}
\sin \theta _{13}\cos ^2\theta _{13}\sin \delta _\text{CP},
\end{equation}
in the PDG parametrization in Ref.~\cite{Workman:2022ynf}.
The $\delta _\text{CP}$ is also determined by one of the absolute values for PMNS mixing matrix elements:
\begin{align}
\left |U_{\tau 1}\right |^2=&
\sin ^2\theta _{12}\sin ^2\theta _{23}+\cos ^2\theta _{12}\cos ^2\theta _{23}\sin ^2\theta _{13}
\nonumber \\
&-2\sin \theta _{12}\sin \theta _{23}\cos \theta _{12}\cos \theta _{23}\sin \theta _{13}
\cos \delta _\text{CP}. 
\end{align} 
Then, we can determine the $\delta _\text{CP}$.
In Fig.~\ref{fig:23_cp}, we show the allowed region for the $\sin ^2\theta _{23}$ and $\delta _\text{CP}$
within $3\sigma $ standard deviation of the Table~\ref{table:NuFIT_data}.
The gray area is outside of  $3\sigma$ standard deviation of $\delta _\text{CP}$ for NuFIT 5.1 data 
in Ref.~\cite{Esteban:2020cvm,Gonzalez-Garcia:2021dve}.
The relation between $\sin^2\theta_{23}$ and Dirac CP phase $\delta_\text{CP}$ has strong correlation.
Especially, in $\sin^2\theta_{23}\in[0.41,0.52]$, this relation has one to one correspondence 
because the complex phase comes from one Yukawa coupling phase $\phi_{DA}$.
%Especially, the Dirac CP phase and lepton mixing angle $\theta _{23}$ are strongly correlated. 
Then if the $\theta _{23}$ is more precise measured 
by the future neutrino oscillation experiments, 
the Dirac CP phase is more precise predicted, and vice versa. 
%Then if experimental restrictions of $\theta_{23}$ or $\delta_\text{CP}$ become stronger 
%around this region, we can strongly predict $\delta_\text{CP}$ or $\theta_{23}$.

%The parametrization in Eq.~\eqref{eq:PMNS_matrix} is the case of the Dirac neutrino. 
%If neutrinos are Majorana particles, 
%If it rotate Majorana field by a phase, this phase appear mass matrix.
%there are three CP phases such as one Dirac CP phase $\delta_\text{CP}$ and two Majorana phases $\eta_1,\  \eta_2$ when we consider three generations. % and Majorana neutrino.

We can  diagonalize the $m_\nu m_\nu^\dag$ in Eq.~\eqref{eq:mass_matrix_left_neutrino} by using the unitary matrix $U_\nu$.
We can also diagonalize the complex symmetric matrix $m_\nu$ by using $U_\nu$ as follows:
%The unitary matrix $U_\nu$ which we obtained from Eq.~\eqref{eq:mass_matrix_left_neutrino} has complex phases of mass eigenstates:
\begin{equation}\label{eq:diag_neutrino_mass}
  U_{\nu}^\dag m_{\nu} U_{\nu}^{*} = \mathrm{diag}(m_1\mathrm{e}^{i\zeta_1},m_2\mathrm{e}^{i\zeta_2},m_3\mathrm{e}^{i\zeta_3}).
\end{equation}
In order to remove these phases from mass diagonal matrix, we need to multiply phase diagonal matrix $P_\nu=\mathrm{diag}(\mathrm{e}^{i\zeta_1/2},\mathrm{e}^{i\zeta_2/2},\mathrm{e}^{i\zeta_3/2})$ on both sides of Eq~\eqref{eq:diag_neutrino_mass},
\begin{equation}
  {P_\nu}^\dag({U_{\nu}}^\dag m_{\nu} {U_{\nu}}^{*}){P_\nu}^{*} = \mathrm{diag}(m_1,m_2,m_3).
\end{equation}
Then, the unitary matrix $U_{\nu}P_{\nu}$ makes the mass matrix real diagonal.
Similarly, we diagonalize $M_\ell M_\ell^\dag$ in Eq~\eqref{eq:MldaggerMl} by using the unitary matrix $U_\ell$.
Therefore we can calculate the PMNS matrix in our model as follows: 
\begin{align} \label{eq:PMNS_model}
	U_{\text{PMNS}}^{\text{mod}} =& {U_\ell}^\dag U_{\nu} P_{\nu} \notag \\
	=&
	\begin{pmatrix}
    		U_{e1}^{\text{mod}} & U_{e2}^{\text{mod}} & U_{e3}^{\text{mod}} \\
		U_{\mu1}^{\text{mod}} & U_{\mu2}^{\text{mod}} & U_{\mu3}^{\text{mod}} \\
		U_{\tau1}^{\text{mod}} & U_{\tau2}^{\text{mod}} & U_{\tau3}^{\text{mod}}
  	\end{pmatrix}.
\end{align}
%Therefore the full parametrization of PMNS matrix when there are three generation Majorana neutrinos is written as~\cite{Workman:2022ynf}
%\begin{equation}\label{eq:new_pmns}
%  U_{\mathrm{PMNS}} = {U_\ell}^\dag U_{\nu} P_{\nu}
%\end{equation}
%\begin{equation*}
%  =
%  \begin{pmatrix}
%    c_{12}c_{13} & s_{12}c_{13} & s_{13}\mathrm{e}^{-i\delta_{\mathrm{CP}}} \\
%    -s_{12}c_{23}-c_{12}s_{23}s_{13}\mathrm{e}^{i\delta_{\mathrm{CP}}} & c_{12}c_{23}-s_{12}s_{23}s_{13}\mathrm{e}^{i\delta_{\mathrm{CP}}} & s_{23}c_{13} \\
%    s_{12}s_{23}-c_{12}c_{23}s_{13}\mathrm{e}^{i\delta_{\mathrm{CP}}} & -c_{12}s_{23}-s_{12}c_{23}s_{13}\mathrm{e}^{i\delta_{\mathrm{CP}}} & c_{23}c_{13}
% \end{pmatrix}
%  \begin{pmatrix}
%   \mathrm{e}^{i\eta_1}&&\\
%    &\mathrm{e}^{i\eta_2}&\\
%    &&1
%  \end{pmatrix},
%\end{equation*}
%where $\eta_1=\frac12(\zeta_1 - \zeta_3)$ and $\eta_2=\frac12(\zeta_2 - \zeta_3)$. 
The Majorana phases $\eta _1$ and $\eta _2$ are determined by using PMNS matrix in Eq.~\eqref{eq:PMNS_model} as follows:
\begin{equation}
\eta _1=\text{arg}\left [
\frac{U_{e1}^{\text{mod}} U_{e3}^{\text{mod}*}}{\cos \theta _{12}\cos \theta _{13}\sin \theta _{13}e^{i\delta _\text{CP}}}\right ], 
\quad 
\eta _2=\text{arg}\left [
\frac{U_{e2}^{\text{mod}} U_{e3}^{\text{mod}*}}{\sin \theta _{12}\cos \theta _{13}\sin \theta _{13}e^{i\delta _\text{CP}}}\right ].
\end{equation}
\begin{figure}[t]
  \begin{minipage}{0.5\hsize}
    \begin{center}
    \includegraphics[width=75mm]{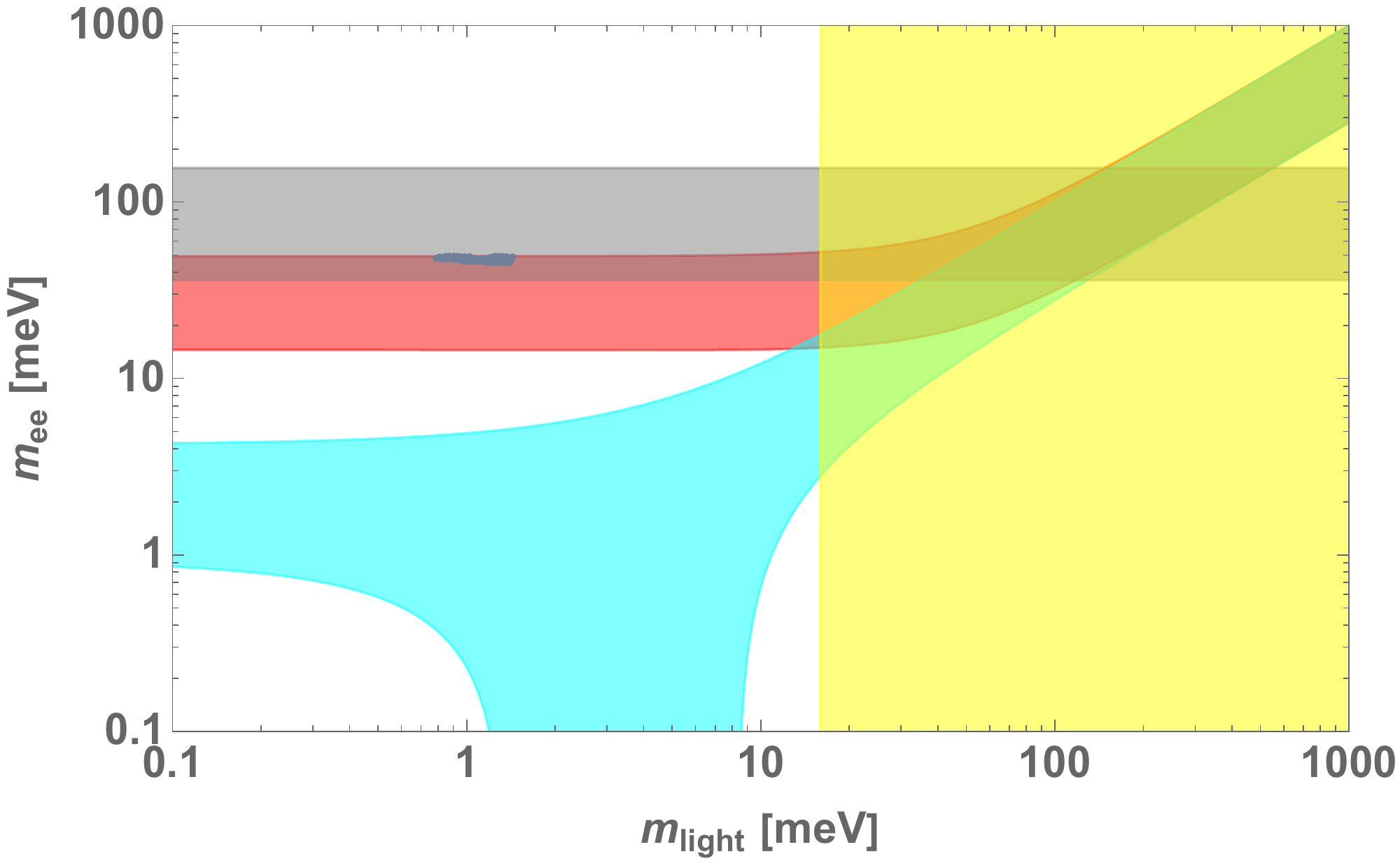}
    \subcaption{}
    \label{fig:mee_mlight}
    \end{center}
  \end{minipage}
  \begin{minipage}{0.5\hsize}
    \begin{center}
    \includegraphics[width=75mm]{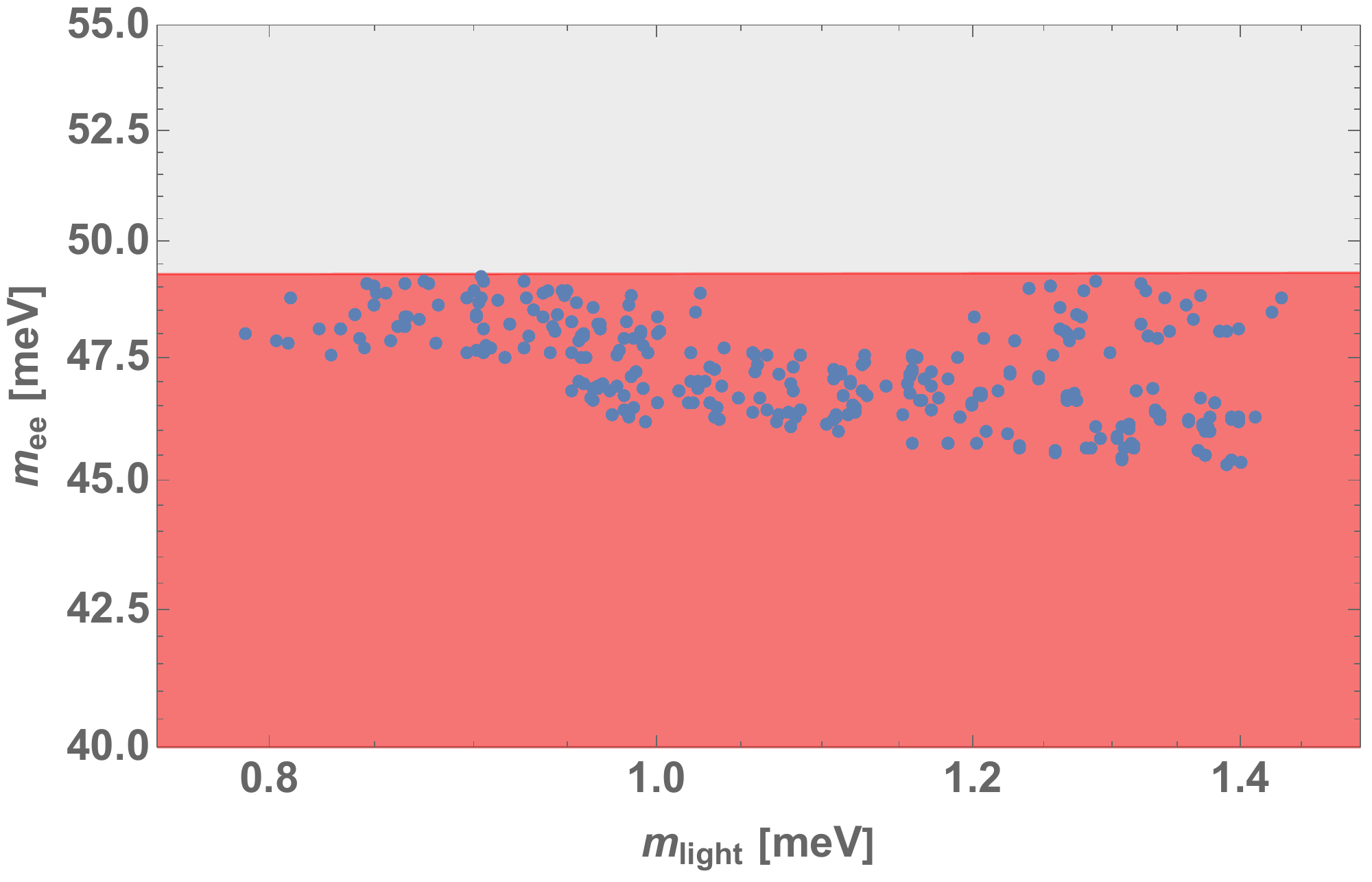}
    \subcaption{}
    \label{fig:expand}
    \end{center}
  \end{minipage}
  
	\begin{center}
	 	\includegraphics[width=75mm]{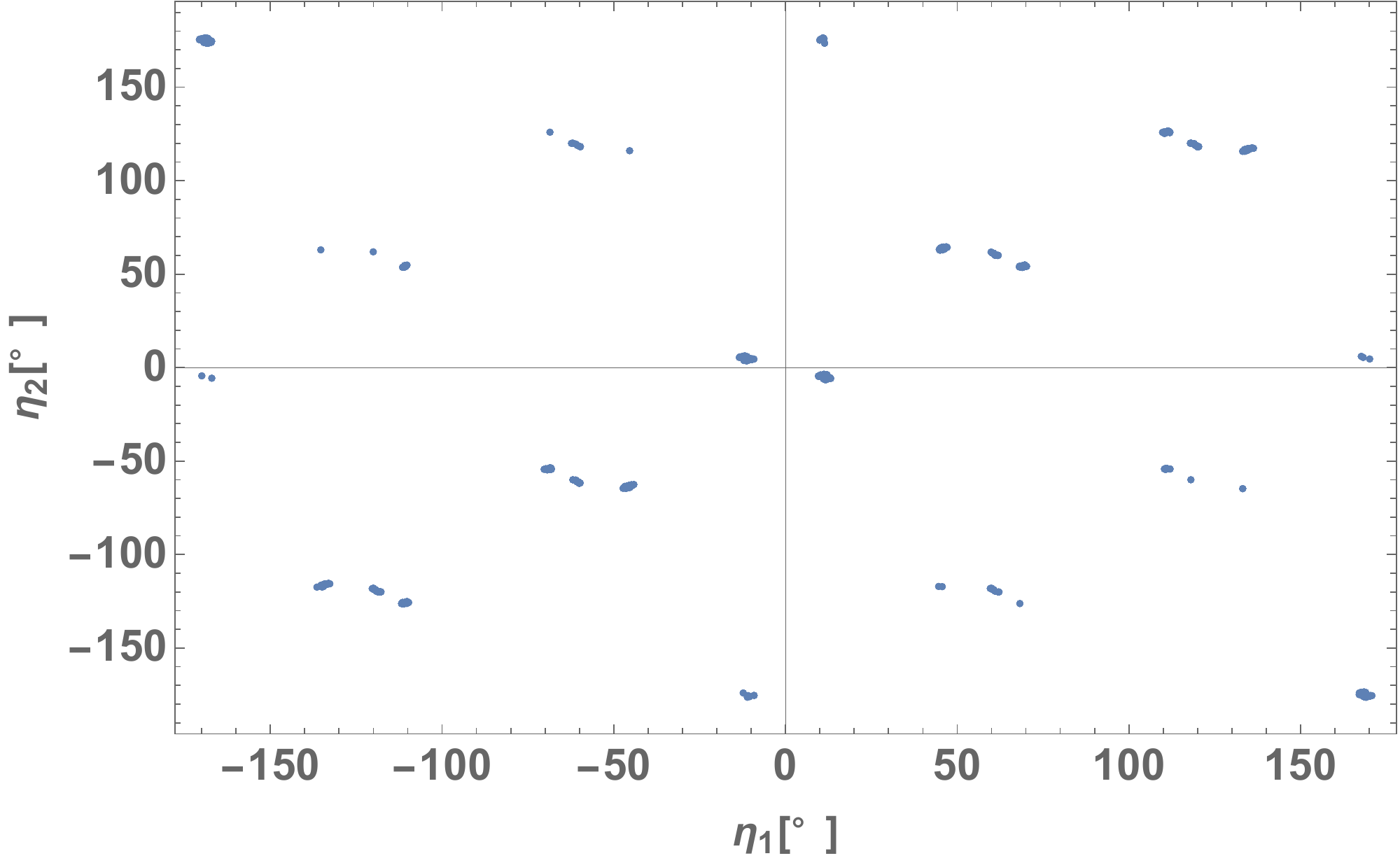}
    	\subcaption{}
    	\label{fig:majorana_phase}
	\end{center}

  \caption{(a)~The relation between the effective mass for $0\nu\beta\beta$ decay 
    $m_{ee}$ and lightest neutrino mass.
    The red area and blue area are model independent analyses for the inverted ordering 
    and normal ordering of the neutrino mass hierarchies, respectively.
    The gray area is upper limit on the effective mass of 36~-~156 [meV] 
    at $90\%$ C.L. in Ref.~\cite{KamLAND-Zen:2022tow}.
    The yellow area is upper limit on the lightest neutrino mass $m_3\simeq 15.9$ [meV] 
    which is estimated in Ref.~\cite{Planck:2018vyg}.
  (b)~The enlargement of figure~(a).
  (c)~The relation between Majorana phases $\eta_1$ and $\eta_2$.}
\end{figure}
The 0$\nu \beta \beta $ decay is determined by the 
magnitude of the lightest neutrino mass and neutrino mass ordering.
In Figs.~\ref{fig:mee_mlight} and \ref{fig:expand} , we show the prediction of %lightest neutrino mass $m_3$ and 
the effective mass $m_{ee}$ for the $0\nu \beta \beta $ decay as 
\begin{equation}
\label{eq:neutrino_effective_mass}
  m_{ee}
  =\left| \sum_{i=1}^{3}m_{i}{U_{ei}^{\text{mod}}}^2 \right|
  =\left| m_1 {U_{e1}^{\text{mod}}}^2 + m_2 {U_{e2}^{\text{mod}}}^2 + m_3 {U_{e3}^{\text{mod}}}^2  \right|.
\end{equation}
Note that only inverted ordering is acceptable then the lightest neutrino mass is $m_3$.
The lightest neutrino mass $m_3$ appears in $0.789$~-~$1.43~[\mathrm{meV}]$ 
which is the within the Planck data~\cite{Planck:2018vyg} and 
the effective mass takes very restricted region $m_{ee}\simeq 47.1~[\mathrm{meV}]$.
This value is in upper limit on the effective mass of 36~-~156 [meV] 
at $90\%$ C.L. in Ref.~\cite{KamLAND-Zen:2022tow}. 
Then this model can be verified in the near future neutrino experiments, e.g. 
the KamLAND-Zen~\cite{KamLAND-Zen:2016pfg,KamLAND-Zen:2022tow}, 
GERDA\cite{GERDA:2013vls,GERDA:2018pmc}, and 
CUORE~\cite{CUORE:2017tlq,CUORE:2019yfd} experiments. 
In Fig.~\ref{fig:majorana_phase}, we show the relation among Majorana phases $\eta _1$ and $\eta _2$. 
Since the phase of our model parameter is only $\phi _{DA}$, then the Majorana phases are 
strongly correlated.

%\newpage 

\section{Summary and Discussions}
\label{sec:Summary}
We have proposed the simple non-SUSY lepton flavor model with $A_4$ symmetry.
The $A_4$ group is a minimal one which includes triplet irreducible representation.
We have introduced three Higgs doublets which are assigned as triplet of the $A_4$ symmetry.
It is natural that there are three generations as same as the SM fermions.
First, we have analysed the potential and we have got the VEV for the local minimum.
Next, we have presented our $A_4$ model. The left-handed lepton doublets are
assigned to triplet and the right-handed leptons are assigned to different singlets of 
the $A_4$ symmetry, respectively.
We have introduced the right-handed Majorana neutrinos which are assigned to triplet of 
the $A_4$ symmetry.
In our model, the right-handed Majorana neutrino mass matrix has a simple flavor structure.
On the other hand, the Dirac neutrino mass matrix has symmetric and anti-symmetric
Yukawa couplings for $A_4$ symmetry.
By using the type-I seesaw mechanism, we have obtained the left-handed 
Majorana neutrino mass matrix.
After diagonalizing the charged lepton and left-handed Majorana neutrino mass matrices,
we have got the PMNS mixing matrix. 
In our numerical analyses, we have used the NuFIT 5.1 data.
%%%%%%%%%   20240910 Takei added %%%%%%%%%%%%%%%%
We found that only inverted ordering is acceptable and we could not find the solutions 
for normal ordering in the neutrino mass hierarchy.
%and normal ordering is rejected.
We have obtained relevant relations for mixing angles and neutrino effective mass $m_{ee}$ 
as a function of the lightest neutrino mass.
Especially, the Dirac CP phase and lepton mixing angle $\theta _{23}$ are strongly correlated. 
%Especially, the relation between $\theta_{23}$ and Dirac CP phase $\delta_\text{CP}$ 
%have strong prediction. 
If the $\theta _{23}$ is more precise measured, 
the Dirac CP phase is more precise predicted, and vice versa. 
In this model, the effective mass for the $0\nu \beta \beta $ decay can be predicted as 
$m_{ee}\simeq 47.1~[\mathrm{meV}]$ and the lightest neutrino mass can be also predicted as 
$m_3\simeq 0.789$~-~$1.43~[\mathrm{meV}]$. 
%which is can be verified in the near future. 
It is testable for near future neutrino experiments. 

The flavor symmetry also apply to the quark sector.
In the same assignments as the charged leptons, the elements of the mass matrices are real. 
Then, we take quarks different assignments, e.g. left-handed quark doublets are assigned to the 
different singlets and right-handed up and down-type quarks are assigned to the triplets of 
the $A_4$ symmetry, respectively. The more details are in the future work.
In our model, the right-handed Majorana mass matrix is very simple and masses are degenerate because the right-handed Majorana neutrino is $A_4$ triplet. 
Then, we cannot apply to the leptogenesis.
Fortunately, there are three Higgs doublets, we can discuss the electroweak baryogenesis which are also 
in the future work.

%-------- acknowledgement -------%
\vspace{1cm}
\noindent
{\large \bf Acknowledgement}
\vspace{1mm}

We thank Y. Kawamura, Y. Matsuo, H. Shimoji, S. Takahashi, and A. Yuu for useful discussions.

%\newpage 
%---------------------------------------------------------%
%--------------- Appendix --------------------------------%
%---------------------------------------------------------%
\appendix
\section*{Appendix}

\section{Multiplication rule of $A_4$ group}
\label{sec:multiplication-rule}
We show the multiplication rule of the $A_4$ triplets as follows:
\begin{align}
\label{eq:multiplication-rule}
\begin{pmatrix}
a_1\\
a_2\\
a_3
\end{pmatrix}_{\bf 3}
\otimes
\begin{pmatrix}
b_1\\
b_2\\
b_3
\end{pmatrix}_{\bf 3}
&=\left (a_1b_1+a_2b_3+a_3b_2\right )_{\bf 1}
\oplus \left (a_3b_3+a_1b_2+a_2b_1\right )_{{\bf 1}'} \nonumber \\
& \oplus \left (a_2b_2+a_1b_3+a_3b_1\right )_{{\bf 1}''} \nonumber \\
&\oplus \frac13
\begin{pmatrix}
2a_1b_1-a_2b_3-a_3b_2 \\
2a_3b_3-a_1b_2-a_2b_1 \\
2a_2b_2-a_3b_1-a_1b_3
\end{pmatrix}_{{\bf 3}}
\oplus \frac12
\begin{pmatrix}
a_2b_3-a_3b_2 \\
a_1b_2-a_2b_1 \\
a_3b_1-a_1b_3
\end{pmatrix}_{{\bf 3}\  } \ .
%%
%{\bf 1} \otimes {\bf 1} = {\bf 1} \ , \qquad &
%{\bf 1'} \otimes {\bf 1'} = {\bf 1''} \ , \qquad
%{\bf 1''} \otimes {\bf 1''} = {\bf 1'} \ , \qquad
%{\bf 1'} \otimes {\bf 1''} = {\bf 1} \  .
\end{align}
More details are shown in the review~\cite{Ishimori:2010au,Ishimori:2012zz}.
%------------------------------------------------------------------------------%
%--------------------------    References    ----------------------------------%
%------------------------------------------------------------------------------%
%\newpage

\end{document}